\documentclass[10pt,letterpaper]{article}
\usepackage{opex3}
 \usepackage{subfigure}

\begin{document}

\title{The noise-limited-resolution for stimulated emission depletion microscopy of diffusing particles}

\author{Chris J. Lee$^{1,2,*}$ and Klaus J. Boller$^{1}$}

\address{$^{1}$Laser Physics \& Nonlinear Optics Group, Faculty of Science and Technology, MESA+ Research Institute for Nanotechnology, University of Twente, P. O. Box 217, Enschede 7500AE, Netherlands\\
$^{2}$FOM Institute DIFFER, Edisonbaan 14, 3439 MN Nieuwegein, The Netherlands}

\email{c.j.lee@differ.nl} 

\begin{abstract}
With recent developments in microscopy, such as stimulated emission depletion (STED) microscopy, far-field imaging at resolutions better than the diffraction limit is now a commercially available technique. Here, we show that, in the special case of a diffusive regime, the noise-limited resolution of STED imaging is independent of the saturation intensity of the fluorescent label. Thermal motion limits the signal integration time, which, for a given excited-state lifetime, limits the total number of photons available for detection.
\end{abstract}
\ocis{(110.0180) Microscopy; (180.2520) Fluorescence microscopy; (110.0110) Imaging systems; (110.4280) Noise in imaging systems.}

\section{Introduction}
\label{sec:intro}
Modern microscopy is one of a number of techniques that has driven the modern biology revolution, revealing important physiological details at systemic, cellular and sub-cellular levels. One of the key limitations of optical microscopy is the diffraction limit, which hides details on scales smaller than the wavelength of the illuminating light. The drive to overcome this limit has led to several notable techniques, such as STED microscopy and stochastic optical reconstruction microscopy (STORM). These techniques are able to achieve quite astonishing resolutions. Examples include imaging frozen cells with a lateral 20-30~nm resolution using STORM~\cite{Huang:2008p4221}, live cell imaging with a lateral resolution of 60~nm using STED~\cite{Hein:2008p4237} and solid state imaging of nitrogen vacancy color centers in diamond with a resolution of 6~nm (and a centroid positional resolution of 0.1~nm)~\cite{Rittweger:2009p8335}.

STED microscopy simultaneously illuminates the sample with two laser beams, both focused to diffraction-limited spot-sizes. One beam is Gaussian shaped and tuned to the absorption band of the fluorescent labels. The second, called the STED laser, possesses a node at the centre of the beam and is tuned to the emission wavelength of the fluorescent labels. The STED laser stimulates the emission of excited fluorescent molecular labels, depleting the excited state population outside of the small volume located at the node of the STED laser beam. As a result, spontaneous emission only occurs from within this smaller volume, which is defined by where the STED laser intensity is lower than the saturation intensity of the label. This working principle suggests that the volume would be reduced to arbitrarily small values simply by increasing the intensity of the STED laser, which, in turn, implies that there is no limit to the achievable spatial resolution.

This leads to the question of what might limit the resolution limit of these new imaging techniques. Current research has focused on how the resolution scales with STED laser intensity~\cite{Harke:2008p7624} and how STED signal levels depend on the details of the label and illumination regime~\cite{Leutenegger:2010p15735}. The possibility of improving the optical resolution of an imaging system through the use of classical and non-classical light sources has been discussed~\cite{Kolobov:2000wy, Beskrovnyy:2005dl}. To our knowledge, no previous work has investigated where the ultimate limits of STED are, because, although the diffraction limit no longer applies, fundamental sources of noise, such as shot noise and thermal motion must, at some point, play a role in limiting the resolution of all sub-diffraction-limited resolution microscopy techniques. With the advent of nano-diamond based fluorophores~\cite{Hui:2010cwa} and the use of STED microscopy for solid-state physics studies~\cite{Wildanger:2011hya}, the ultimate resolution of STED gains importance, because other limiting factors, such as bleaching, are eliminated.

In this paper, we analyze the behavior of the signal-to-noise ratio of STED microscopy. In particular we examine how the Brownian motion of the sample and photon shot noise influence the resolution of STED microscopy. In general, STED imaging is performed by raster scanning the STED laser over the field of view. At each location, signal is acquired for a particular dwell time---a time that is limited by the shot noise of the system. During this time, objects can diffuse into and out of the volume corresponding to the current location, blurring the image. It is, therefore, necessary for the signal acquisition time to be shorter than the diffusion time of the features that one wishes to resolve. Once these aspects of fundamental noise are taken into account, it is found, surprisingly, that the noise-limited resolution is independent of the fluorophore's saturation intensity, but, rather depends on the excited state lifetime of the fluorophore, the density of fluorophores and the type of diffusive regime of the object being imaged.
 
\section{Theory}
\label{sec:theory}
For a STED microscope, the lateral resolution is given by $dr = \frac{\lambda}{2\textrm{NA}}\sqrt{\frac{I_{s}}{I}}$ ($I_{s}/I<$1)~\cite{Hein:2008p4237}, where $I$ is the peak intensity of the STED laser, and $I_{s}$ is the saturation intensity of the fluorophore. The volume from which light is collected from a standard confocal microscope is $V_{0} = \frac{\pi\lambda^{3}}{4\textrm{NA}^{3}}$, where $\lambda$ is the wavelength of the excitation source, and NA is the numerical aperture of the microscope objective. For a STED microscope, the increase in resolution leads to a reduced volume, $V$, given by:
	\begin{equation}
	\label{STED volume}
	V = \frac{V_{0}I_{s}^{\frac{3}{2}}}{I^{\frac{3}{2}}}
	\end{equation}
where we have approximated the axial resolution as $2dr$. 

To obtain the photon flux from $V$, we consider a sample that has a density $\rho_{e}$ of fluorophores that have a cycling time---the time it takes for the fluorophore, once excited in the absence of the STED light field, to return to the ground state---$\tau_{c}$, and a saturation intensity $I_{s}$. 
Assuming that we drive the fluorophore in the saturation regime (this corresponds to the case where the fluorophore is emitting the maximum number of photons per unit time), then the number of photons per unit time will be given by
	\begin{equation}
	\label{photon flux}
	\Phi_{p} = \frac{\rho_{e}V}{\tau_{c}}
	\end{equation}
As expected, the photon flux reduces with increasing STED laser intensity and increases if a fluorophore with a reduced excited state lifetime is chosen. This photon flux determines the maximum signal to noise ratio, $R$, that can be achieved in detection. Assuming Poisson statistics (the fluctuations in $\Phi_{p}$ are proportional to $\sqrt{\Phi_{p}}$), the detector current signal to noise ration is given by:
	\begin{eqnarray}
	\label{SNR}
	R =\int_{0}^{t} \frac{i}{\bar{i}_{rms}}dt\\
	\label{detector current}
	i = e\Theta\Phi_{p}
	\end{eqnarray}
where $i$ is the detector current, $\bar{i}_{rms}$ is the noise current, $e$ is the electron charge, $t$ is the measurement time, and $\Theta = (\textrm{NA})T\eta$ is the overall efficiency of collecting and detecting a photon ($T$ is the total transmission of the optical system and $\eta$ is the quantum efficiency of the detector). In a measurement time interval of $t$, the signal-to-noise ratio is found by evaluating eq.~\ref{SNR}:
	\begin{equation}
	\label{STED signal to noise}
	R = \sqrt{\frac{\Theta\rho_{e}Vt}{\tau_{c}}}
	\end{equation}
	
In principle $R>$1 is always obtained for a sufficiently large $t$. But, in practice, $t$ is limited. This is illustrated by considering the case of particles (with fluorophores attached) that are subject to diffusion. In this case, the maximum measurement time is given by the time for which a particle remains within the observation volume. Given a diffusion constant $D$ (which is size dependent), in time, $t$, a particle will be displaced by an average distance $\delta r$
	\begin{equation}
	\label{diffusion displacement}
	\delta r = \sqrt{Dt^{\alpha}} 
	\end{equation}
where $\alpha$ (which is size dependent) is an empirically derived value: $\alpha$=1 corresponds to normal diffusion, while $\alpha<$1 and $\alpha>$1 correspond to sub-diffusive and super-diffusive regimes, both of which are important for biological processes~\cite{Caspi:2000vd, Caspi:2002eba}. The maximum observation time is the time it takes for a particle to diffuse out of the observation volume.
	
Solving eq.~\ref{diffusion displacement} for $t$, substituting into eq.~\ref{STED signal to noise}, and setting $\delta r = dr$, provides a relationship between the size of the observation volume, the observation time, and the photon flux. 
	\begin{equation}
	\label{vol sign to noise}
	R = \sqrt{\frac{\Theta\rho_{e}Vdr^{\frac{2}{\alpha}}}{D^{\frac{1}{\alpha}}\tau_{c}}}
	\end{equation}

By requiring $R$=1 in eq.~\ref{vol sign to noise}, the minimum reduced volume, as a function of diffusion properties, is obtained. From eq.~\ref{STED volume} and $dr = \frac{\lambda}{2\textrm{NA}}\sqrt{\frac{I_{s}}{I}}$, the lateral resolution as a function of diffusion properties is found:
	\begin{equation}
	\label{STED resolution}
	dr_{s} = \frac{1}{2^{\frac{2\alpha-1}{3\alpha+2}}}\left[\frac{\tau_{c}}{\pi\Theta\rho_{e}}\right]^{\frac{\alpha}{3\alpha + 2}}D^{\frac{1}{3\alpha+2}}
	\end{equation}
where we have also made a substitution for $V_{0}$. Note that eq.~\ref{STED resolution} is still dependent on the wavelength of the lasers and the numerical aperture of the system via the collection efficiency $\Theta$. 

Eq.~\ref{STED resolution} does not give the resolution as an explicit function of the saturation intensity of the fluorophore, however, in understanding the influence of saturation intensity and diffusion on resolution, it is more convenient to include $I_{s}$ explicitly. Eq.~\ref{STED resolution} shows that the resolution is a function of the cycling time of the fluorophore and the fluorophore density (e.g., more fluorophores attached to a single particle). Although it seems that the resolution is independent of the saturation intensity, the cycling time and the saturation intensity are related to each other. The saturation intensity is given by~\cite{Verdeyen:wa}
	\begin{equation}
	\label{sat intensity equation}
	I_{s} = \frac{8\pi\sqrt{\pi} hcn^{2}}{\lambda^{3}}\frac{\tau_{e,j}}{\tau_{e}^{2}}
	\end{equation}
where $\lambda$ is the wavelength of the STED laser, $n$ is the refractive index, $\tau_{e,j}$ is the lifetime of the $e\rightarrow j$ transition that the STED laser uses to stimulate emission, and $1/\tau_{e} = \sum_{k\neq j}1/\tau_{e,k}$ is the total lifetime of the excited state of the fluorophore. For simplicity, we have also assumed that the STED laser is tuned to the peak of the stimulated emission cross-section. The fluorophore cycling time is $\tau_{c} = \tau_{e,j} + \tau_{\rightarrow e} + \tau_{\rightarrow g}$, where $\tau_{\rightarrow e}$ is the time taken to enter the excited state (after excitation to some upper state by a laser) and $\tau_{\rightarrow g}$ is the time taken to return to the ground state (perhaps via intermediate states) after fluorescing. Assuming that $\tau_{c}\gg\tau_{\rightarrow e}, \tau_{\rightarrow g}$ and substituting for the lifetime in Eq.~\ref{STED resolution} gives
	\begin{equation}
	\label{STED resolution Is dependence}
	d_{rs}\approx \frac{1}{2^{\frac{2\alpha-1}{3\alpha+2}}}\left[\frac{8\sqrt{\pi}n^{2}hc}{\Theta\rho_{e}\lambda^{3}I_{s}}\right]^{\frac{\alpha}{3\alpha + 2}}D^{\frac{1}{3\alpha+2}} 
	\end{equation}
It can be seen that the noise-limited resolution of STED depends only weakly on the saturation intensity: $\propto (1/I_{s})^{\frac{\alpha}{3\alpha + 2}}$. For a 20~nm particle in the strongly sub-diffusive regime ($\alpha$~=~1/3), even if the saturation intensities vary between 0.1--100~MW/cm$^{2}$, the noise-limited resolution will only change by a factor of two. The same particle in the super-diffusive regime ($\alpha$~=~2) and the same range of saturation intensity results in the noise-limited resolution changing by a factor of 5. This indicates that, ultimately, the diffusive regime of the sample will be the dominant factor in determining the noise-limited STED resolution.

\section{Discussion}
\label{sec:discussion}

\begin{figure}
\label{overview}
\center
\includegraphics[width=6cm]{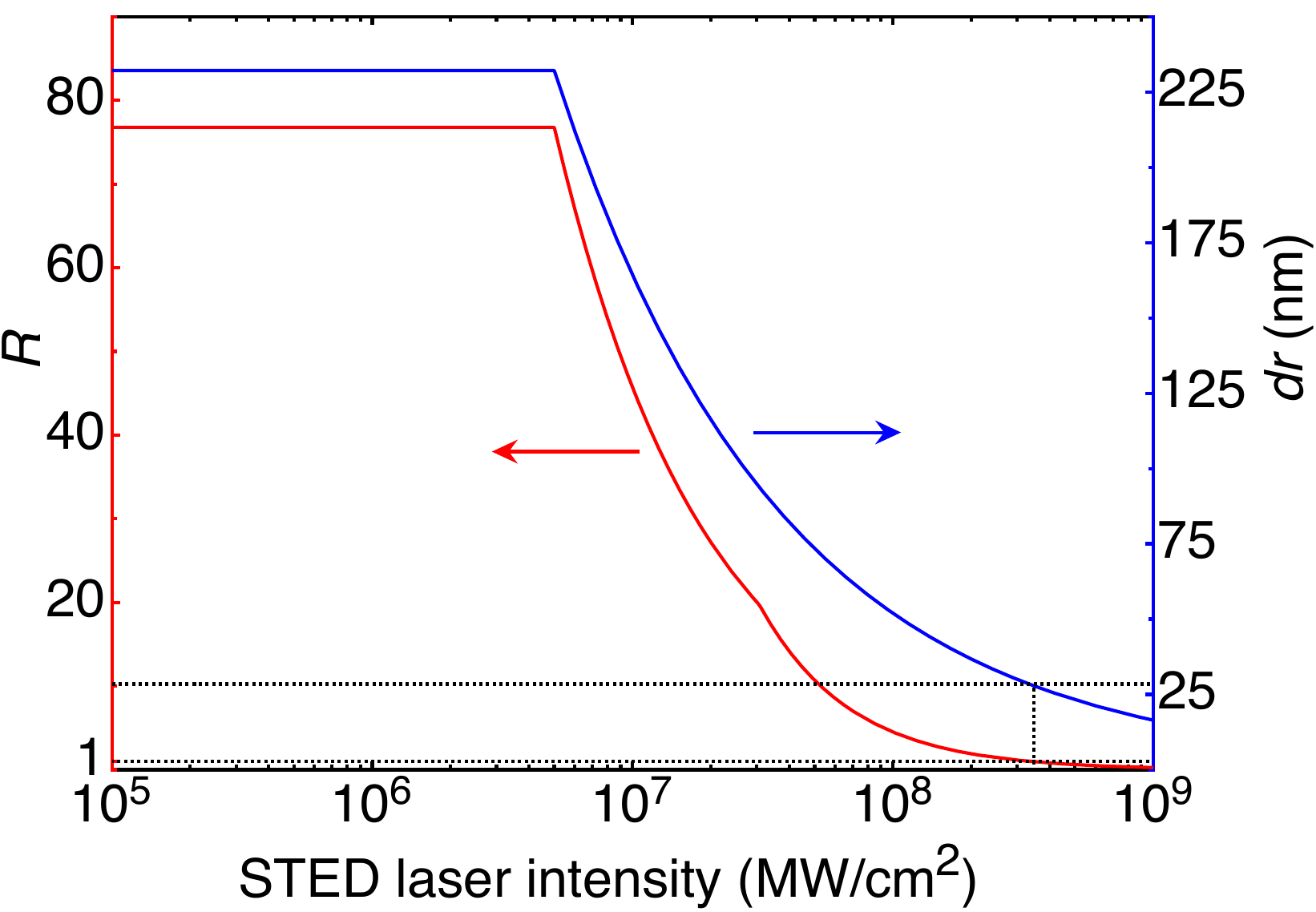}
\caption{Signal-to-noise ratio (red) and lateral resolution (blue) as a function of STED laser intensity. Note that when the STED laser intensity falls below that of $I_{s}$, the stimulated emission rate becomes negligible and the resolution is given by the diffraction-limit.}
\end{figure}

The limits of STED resolution can be estimated by taking appropriate values for the parameters in eq.~\ref{STED resolution}. We consider the case of normal ($\alpha$=1) diffusion of 20~nm objects that have fluorophores ($I_{s}$~=~5~MW/cm$^{2}$~\cite{Rittweger:2009p8335}, $\tau_{2}$~=~4~ns) attached at an average density of 10$^{14}$~cm$^{-3}$ and sitting in a room temperature fluid with the viscosity of water, 1~mPa.s. The microscope is considered to have a NA of 1.4, $\Theta$~=~0.75, and the wavelength is 650~nm. Under these conditions, the best resolution is 28.5~nm, which is obtained for $I/I_{s}$~=~66.7, a diffusion-limited dwell time of 0.4~$\mu$s, and a signal-to-noise ratio of unity (see Fig.~\ref{overview}). The resolution calculated here is comparable to that achieved in live-cell imaging with current STED microscopes, providing an indirect validation (the absolute numbers will depend on experimental details). In STED microscopy resolutions are typically 30-70~nm at ratios of $I/I_{s}\sim$200 and dwell times of 50~$\mu$s are typical for live-cell imaging. Qualitatively, the achieved STED resolution is similar to the expected diffusion distance: a 20~nm object can expect to move about 40~nm in 50~$\mu$s, while a 3~nm object (the hydrodynamic radius of green fluorescent protein) will be displaced by 90~nm on average. These numbers suggest that current STED imaging resolution, on samples such as freely diffusing polystyrene beads, may already be limited by diffusion.

There are two other interesting cases: super-diffusion, and sub-diffusion regimes. In this case, the time scale becomes very important. For instance, for rather large particles (3~$\mu$m diameter) inside cells, it has been shown that on short time scales ($t<$~1~s), motion is super-diffusive ($\alpha$~=~3/2), but over long scales, the motion is sub-diffusive (1~$<\alpha\geq$~0.5)---the motion of the particles is bounded by the cell walls~\cite{Caspi:2000vd, Caspi:2002eba}. In the same work, it was shown that smaller beads had less pronounced super-diffusion ($\alpha<$~3/2) and in \cite{Weiss:2004fra}, it was found that crowding results in sub-diffusion. 

\begin{figure}
\label{alpha dependence}
\center
\includegraphics[width=6cm]{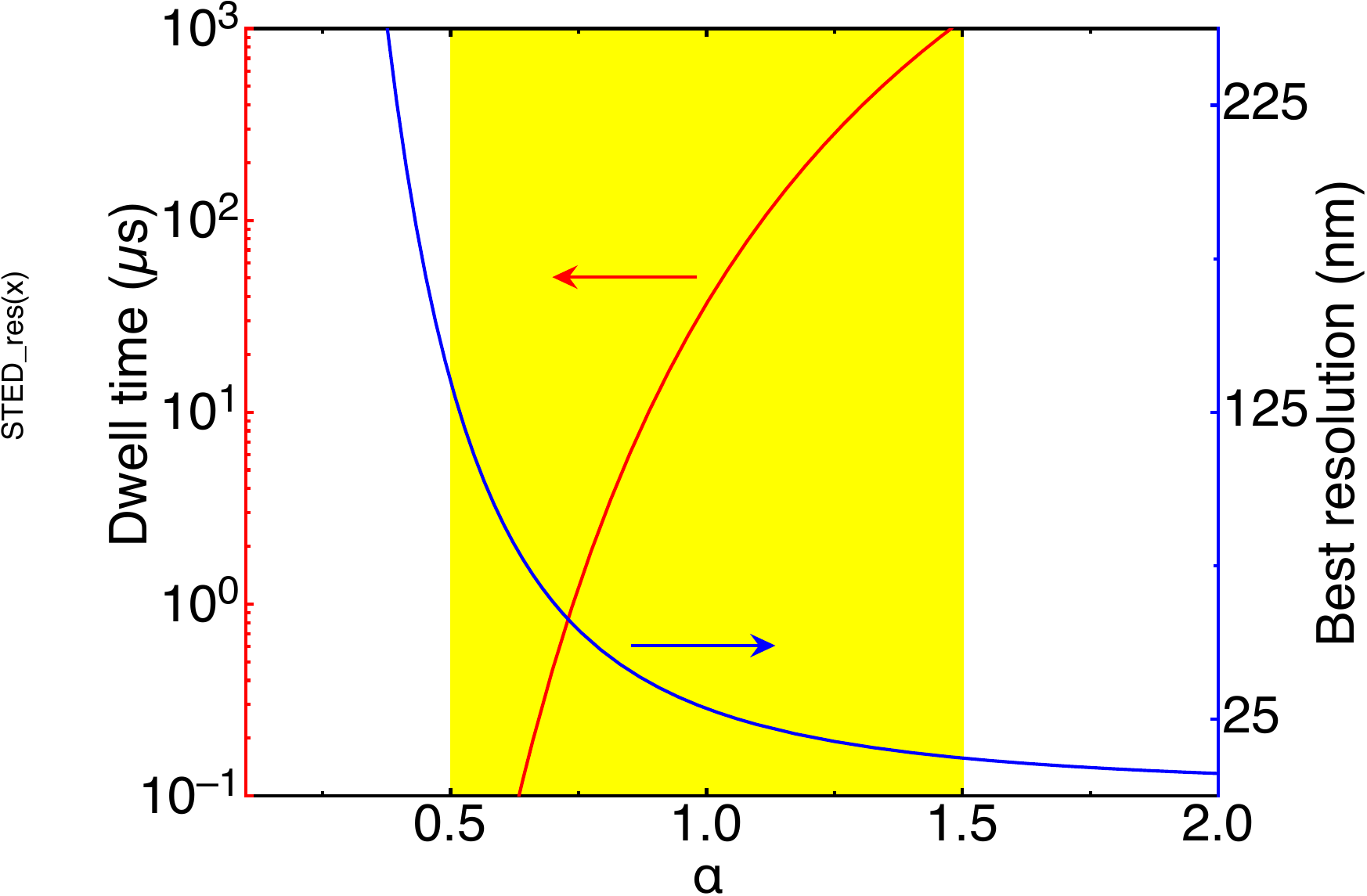}
\caption{Diffusion-limited signal acquisition time (red) and resulting STED resolution (blue) as a function of $\alpha$. The yellow shaded region is the diffusion regime that is of biological relevance.}
\end{figure}

In terms of the obtainable STED resolution, the diffusive regime plays a critical role. Figure~\ref{alpha dependence} shows how the maximum dwell time and resulting resolution depend on the diffusion regime. Clearly, STED imaging is more favorable in the super-diffusive regime, where we, under the conditions described in \cite{Caspi:2002eba} ($\alpha$~=~3/2), calculate that a resolution of 12~nm might be achievable. However, to acquire an image with such resolution requires a dwell time of $\sim$1~ms, which, with current fluorescent labels, is impossible due to bleaching. With the development of nanodiamond-based fluorescent labels, bleaching is expected to be a much less severe limitation. As a result, noise-limited resolution may be obtainable in this regime.

In contrast, the sub-diffusive regime is strongly unfavorable for STED imaging, with the best resolution being 125~nm at $\alpha$~=~0.5, and requires a dwell-time of less than 1~ns. This seemingly counter intuitive result is a consequence of the formalism used to describe diffusion. At $\alpha=$~1 and $t=$~1, a cross over occurs, where as $t$ increases, sub-diffusive behavior becomes super-diffusive, while super-diffusive behavior becomes sub-diffusive. Such cross-over behavior is observed in \cite{Caspi:2002eba}.

\section{Conclusion}
\label{sec:conclusion}
These calculations show that, although the diffraction limit does not limit the resolution of STED microscopy, its resolution is limited by the combination of shot noise and thermal noise. Although the exact resolution limit depends on the excited state lifetime and density of fluorophores, a more important limitation is the diffusive regime of the object being imaged. We show that typical noise-resolution limits are $\sim$30~nm. More interestingly, in the diffusive regime typical of certain cellular features, such as vesicles, it should be possible to image with a resolution of $\sim$12~nm, which is comparable to that achieved with PALM/STORM, though, the time to acquire a single image ($\sim$5 minutes) would either require independent cell motion tracking or that the cell is fixed prior to imaging. We also note that many cell structures do not diffuse within the cell. In this case, the signal acquisition time is not limited by diffusion. Our calculations have focused on fundamental noise, but technical noise should be considered as well. For instance, at a resolution of 6~nm and centroid positional accuracy of 0.1~nm, STED technology has already reached the point where instrumental concerns, such as microscope stage drift must be taken into account. With drifts commonly on the order of 0.1~nm.s$^{-1}$~\cite{NugentGlandorf:2004p8629}, any dynamics extracted from such high resolution images must account for and remove such systemic noise.

\section*{Acknowledgements}
CJL acknowledges funding from: `Controlling photon and plasma induced processes at EUV optical surfaces (CP3E)Õ of the `Stichting voor Fundamenteel Onderzoek der Materie (FOM)Õ, which is financially supported by the `Nederlandse Organisatie voor Wetenschappelijk Onderzoek (NWO)Õ. The CP3E programme is co-financed by Carl Zeiss SMT and ASML.
\end{document}